\documentclass{aastex62}

\graphicspath{{./}{figures/}}
\submitjournal{ApJ}
\shorttitle{Sparse Mapping}
\shortauthors{Aizawa, Kawahara \& Fan}
\usepackage{color}
\definecolor{red}{rgb}{0.8,0.0,0.0}

\usepackage{bm}
\usepackage{amsmath,amsthm,amssymb,cases}
\newcommand{\rev}{\textcolor{black}} 
\newcommand{\revn}{\textcolor{black}} 
\newcommand{\revnn}{\textcolor{black}}

\begin{document}

\title{Global Mapping of an Exo-Earth using Sparse Modeling}

\correspondingauthor{Masataka Aizawa}
\email{aizawa@utap.phys.s.u-tokyo.ac.jp}
\author[0000-0001-8877-4497]{Masataka Aizawa}
\affiliation{Department of Physics, The University of Tokyo, Tokyo 113-0033, Japan}
\author[0000-0003-3309-9134]{Hajime Kawahara}
\affiliation{Department of Earth and Planetary Science, The University of Tokyo, 7-3-1, Hongo, Tokyo, Japan}
\affiliation{Research Center for the Early Universe, 
School of Science, The University of Tokyo, Tokyo 113-0033, Japan}
\author[0000-0002-3041-4680]{Siteng Fan}
\affiliation{Division of Geological and Planetary Sciences, California Institute of Technology, Pasadena, CA 91125, USA}

\begin{abstract} 
We develop a new retrieval scheme for obtaining two-dimensional surface maps of 
exoplanets from scattered light curves. In our scheme, the combination of the 
L1-norm and Total Squared Variation, which is one of the techniques used in 
sparse modeling, is adopted to find the optimal map. We apply the new method 
to simulated \rev{scattered} light curves of the Earth, and find 
that the new method provides a better spatial resolution of the reconstructed 
map than those using Tikhonov regularization. 
We also apply the new method to observed scattered light curves of the Earth 
obtained during the two-year DSCOVR/EPIC observations presented by 
\cite{2019ApJ...882L...1F}. \rev{The method with Tikhonov regularization enables us to resolve
North America, Africa, Eurasia, and Antarctica. In addition to that, the 
sparse modeling identifies South America and Australia, although it fails to find the 
Antarctica maybe due to low observational weights on the poles.} 
Besides, the proposed method is capable of retrieving maps from noise injected light curves of a hypothetical Earth-like exoplanet at \revn{5} pc with noise level \rev{expected from coronagraphic images from} \revnn{a 8-m space telescope}.  We find that the sparse modeling \rev{resolves Australia, Afro-Eurasia, North America, and South America} using 2-year observation with a time interval of one month. Our study shows that the combination of sparse modeling and multi-epoch observation with 1 day \revn{or 5 days} per month can be used to identify main features of an Earth analog in future direct imaging missions  \revnn{ such as the Large UV/Optical/IR Surveyor (LUVOIR)}.

\end{abstract}


\section{Introduction} \label{sec:intro} 
One of the ultimate goals of exoplanetary science is to detect life and 
 characterize habitable environments beyond the Solar System. 
Recent discoveries of exoplanets in the habitable zone (HZ) provide unprecedented 
opportunities to characterize planetary environments that could potentially 
harbor life \citep[e.g][]{2016ApJ...830....1K}.  Beyond the 
detection of planets in the HZ, a promising biomarker on such 
planets in the HZ is the metabolic feature, such as $O_2$, $O_3$, $H_2O$, in 
transmission spectra. Notably, \cite{2019ApJ...887L..14B} have recently 
reported the discovery of water vapor in the atmosphere of an 8 
$M_\oplus$ planet in the HZ (K2-18b). This 
discovery triggers significant motivation for seeking further indirect or 
direct indications of life. 

\rev{In direct imaging, an exoplanet is generally accompanied by a star, whose brightness and proximity 
make identification of planetary light challenging.} In principle, features in scattered light can be 
interpreted as indications of movements of surface inhomogeneity resulting from orbital and rotational motions 
\citep{2001Natur.412..885F}. 
Despite the technical challenges of directly imaging Earth-like 
exoplanets, such a characterization of photometric variations is 
an important probe of exoplanetary surface environments (which may include oceans, land masses, and regions with vegetation) and planetary dynamics \citep{2008ApJ...676.1319P,2009ApJ...700..915C,2009ApJ...700.1428O,2010ApJ...715..866F, 2011ApJ...731...76C,2011ApJ...738..184F}. 

One of the characterization methods based on photometric variations, proposed by \cite{2010ApJ...720.1333K}, involves 
two-dimensional (2D) global mapping of directly imaged planetary
surfaces, using the scattered light curve in the presence of both  
spin and orbital motions. Subsequently, \cite{2011ApJ...739L..62K} 
formulated an inversion method named ``spin--orbit tomography'' 
(SOT) for recovering 2D-surface maps by introducing Tikhonov 
regularization, which enables the direct estimation of 
the surface albedo. The same technique was applied successfully by 
\cite{2012ApJ...755..101F} to 
reconstruct 2D maps even for planets with various obliquities 
and orbital inclinations. Within the framework of SOT, 
planet obliquity is simultaneously inferred through the 
minimization of a loss function. The ability to infer obliquity 
from a light curve was extensively 
investigated by \cite{2016MNRAS.457..926S} in terms of amplitude 
modulation. Also, \cite{2016ApJ...822..112K} revealed the relation 
between the planet spin axis and the frequency modulation of 
light curves. \cite{2018AJ....156..146F} constructed the Bayesian 
framework of 2D mapping and quantified the uncertainty in the 
albedo map and the obliquity. They applied the Gaussian process 
to regulate the inter-pixel variance of the map instead of the 
the Tikhonov regularization. \rev{On the other hand, \cite{2019AJ....157...64L} presented the open code 
{\tt starry} that exploits spherical harmonics for mapping, and \cite{2019arXiv190312182L} 
applied it to the TESS light curves of the Earth. Since TESS is equipped with single bandpass, 
the map recovered by \cite{2019arXiv190312182L} is
largely affected by the cloud reflection. On the other hand, 
\cite{2019ApJ...882L...1F}  recently 
recovered a global land map by exploiting} multi-wavelength 
light curves of the Earth during the two-year DSCOVR/EPIC 
observations \citep{2018AJ....156...26J}. This is a clear 
practical example of how to mitigate the effects of clouds from light curves.

The aim of this study is to improve the spatial 
resolution of an inferred map using sparse modeling. 
\rev{The global mapping is generally a ill-posed problem, so it requires 
regularizations for giving unique solutions. One possible technique is 
Least Absolute Shrinkage and Selection Operator (LASSO), which minimizes the $l_{1}$ norm simultaneously to search for the sparse solutions 
\citep[e.g.][]{1996_tibshirani}. 
Recently, this technique has attracted} increasing attention 
in the field of astronomy, especially in reconstructing images from  interferometric observations \citep[e.g.][]{2014PASJ...66...95H, 2016PASJ...68...45I, 2017ApJ...838....1A, 2017AJ....153..159A, 2018ApJ...858...56K, 2019ApJ...875L...4E}. \rev{For example, \cite{2018ApJ...858...56K} proposed to use the combination of Total Squared variations (TSV) and the $l_{1}$ norm, and they demonstrated that the 
technique enabled us to recover accurate images of black hole shadow from interferometric observations compared with conventional methods.} 

\rev{In this paper, we apply the sparse modeling to mapping 
problem of planetary surfaces from reflectional light in direct-imaging observations.} As fiducial data, we adopted the mock albedo map of the Earth and the real observational data obtained from \revnn{t}he Deep Space Climate Observatory (DSCOVR) by \cite{2019ApJ...882L...1F}. 
For comparison, we solved the mapping problem using 
both Tikhonov regularization and sparse modeling, and 
discuss the difference in the output maps.  

\section{Formulation of the mapping}
\subsection{Forward modeling of reflected light curves of a distant planet}
For mapping the surface of an exoplanet, we estimate the planetary surface albedo $\bm{m}$, discretized to 
$N_{\rm pixel}$ pixels ($ \bm{m} = \{ m_j = m (\theta_j, \phi_j) \}  
$ for $j=1,2,...,N_{\rm pixel} $), where ($\theta_{j}, \phi_{j}$) is the location 
of the $j$th pixel on the sphere's surface. The scattered light 
curve $\bm{d}$ consists of $N_{\rm data}$ points ($ \bm{d} = \{ d_i = d(t_i) \}  
$ for $i=1,2,...,N_{\rm data} $), where $t_i$ is the $i$th time frame. As formulated previously \citep[e.g][]{2010ApJ...720.1333K}, 
the light curve and planetary surface are related via 
a transfer matrix \rev{$\bm{G} = \{ G_{i, j} = G (\theta_{j}, \phi_{j}; \Phi_i, \Theta_i; \zeta, \Theta_{\rm eq}, i_{\rm inc}) \} $} for $i=1,2,...,N_{\rm data} $ and  $j=1,2,...,N_{\rm pixel} $ as  
\begin{equation}
d_{i} = \sum_{j}^{N_{\rm pixel}} G_{i, j} m_{j} + \epsilon_{i}, \label{d_G_m_eq}
\end{equation}
where $\bm{G}$ describes the amount of 
reflection from the planetary surface toward the observer with a given influx from the central star, and $\bm{\epsilon}$ corresponds to the observational uncertainties associated 
with $\bm{d}$. Here, $\Phi_i = \Phi(t_i)$ and $\Theta_i = \Theta (t_i)$ 
describe the phases of the orbital motion and spin rotation, respectively; 
$\zeta$ is the planetary obliquity, \rev{the angle between the planetary orbital axis and 
the planetary spin axis;} $\Theta_{\rm eq}$ is the orbital 
phase at the equinox; \rev{$i_{\rm inc}$ is the planetary orbital inclination, which is defined as the angle between 
the orbital axis and the line-of-sight. In the following manuscript, we assume a circular orbit for simplicity.} 
These geometrical quantities can be inferred from 
the mapping itself \citep[e.g.][]{2010ApJ...720.1333K,2016MNRAS.457..926S,2018AJ....156..146F} 
and also from the frequency modulation \citep{2016ApJ...822..112K}. 
Hence, we fix $\zeta$ and $\Theta_{\rm eq}$ to their true values for simplicity throughout this letter.

\rev{Since we do not have any information in surface types, we assume isotropic (Lambertian) reflection 
for simplicity in this paper. We note that one can also attempt to find signatures of 
non-isotropic reflection from light curves, which might indicate the presence of 
particular surface types, e.g. ocean \citep{2018AJ....156..301L}. } 
Assuming isotropic (Lambertian) reflection from the surface, 
$\bm{G}$ is rewritten as follows:
\[
    G_{i, j} \equiv  
    \begin{cases}
     (e_{\rm S}\cdot e_{\rm R}) (e_{\rm O} \cdot e_{\rm R}) \Delta \omega_{s} & \text{if}\;\; e_{\rm S} \cdot e_{\rm R}>0, e_{\rm O} \cdot e_{\rm R}>0 \\
    0 & \text{otherwise,}
    \end{cases}
\]
where $\Delta \omega_{s}$ is the solid angle subtended by the pixel, and 
$e_{\rm S}, e_{\rm O} $, and $ e_{\rm R}$ are the unit vectors pointing from 
the discretized planetary surface specified by $(\theta_{j}, \phi_{j})$ 
toward the central star, from the surface to the observer, and from the planetary 
center to the surface, respectively. \rev{Their expressions are given by 
\begin{align}
e_{\rm S} &= (\cos(\Theta - \Theta_{\rm eq}), \sin(\Theta - \Theta_{\rm eq}), 0)^{T}, \\
e_{\rm O} &= (\sin i_{\rm inc} \cos \Theta_{\rm eq}, -\sin i_{\rm inc} \sin \Theta_{\rm eq}, \cos i_{\rm inc})^{T}, \label{eq:obs} \\
e_{\rm R} = (\cos(\phi + \Phi)\sin\theta, \cos \zeta &\sin(\phi + \Phi) \sin \theta +  \sin \zeta \cos \theta, -\sin \zeta \sin(\phi + \Phi) \sin \theta +  \cos \zeta \cos \theta)^{T}.
\end{align}
}
The transfer function ${\bf G}$ has two fundamental timescales, associated with the 
spin and orbital periods. Combinations of these timescales 
allow the consideration of different positions on the 
planetary surface at different epochs. The inversion method 
exploiting this property is called “spin--orbit tomography” (SOT) 
\citep[see more detailed discussions in][]{2010ApJ...720.1333K,2011ApJ...739L..62K,2012ApJ...755..101F}.

\subsection{Inverse modeling of reflected light curves of a distant planet using 
regularization terms}

\rev{The mapping problem to infer $m$ from $d$ in Eq (\ref{d_G_m_eq}) generally becomes ill-posed 
for two reasons: (a) There possibly exist invisible faces of planets seen from an observer depending on geometry. (b) Angular resolution for mapping is basically limited by visible and illuminated area for each observational snapshot, so it is not impossible to reconstruct the planetary surface at infinite resolution. The first property can be mitigated by limiting the solution space not to include invisible areas, but the second one is still inevitable. There are several methods to give an unique solution in a ill-posed problem, and one of such techniques is an introduction of regularization terms to a standard chi-squared value in minimization. 
Previously, \cite{2011ApJ...739L..62K} introduced the Tikhonov regularization, or also 
known as ridge regression, to determine the unique solution. }

\rev{Although it successfully gives the unique solution, 
the Tikhonov regularization is not the only choice of regularizations. 
By choosing appropriate regularization parameters, 
one can exploit unique features of planetary surfaces in 
direct-imaging observations; (a) the surface is composed of several major types, possibly with a few dominant species e.g. ocean 
in case of the Earth; (b) the surface is likely to be continuous and smooth; (c) coast lines on the surface, 
on the others, are likely to be sharp, and they divide different surface types clearly. 
In this paper, we attempt to exploit the first two features by introducing 
the L1-norm and Total Squared Variation (TSV), and we evaluate the 
difference between the map recovered by the Tikhonov regularization and 
the new regularization. We also attempt other types of regularization terms including Total Variation, which can exploits the property (c), 
in Appendix \ref{sec:other_reg}, and we find that the combination of 
the L1-norm and TSV is likely to give the best estimation on the map. }

\subsubsection{Modeling with Tikhonov regularization}
To solve $\bm{m}$, \cite{2011ApJ...739L..62K} 
introduced the Tikhonov regularization: 
\begin{align}
Q_{\rm \lambda} &\equiv  \chi^{2} + \lambda^{2}|\bm{m}-\hat{\bm{m}}|^{2}, \label{Eq:tikhonov}
\\
 \chi^{2} &= \sum_{i=1}^{N_{\rm data}} \frac{(d_{i} - \sum_{j=1}^{N_{\rm pixel}}G_{i,j} m_{j})^{2}}{\sigma_{i}^{2}}, 
\end{align}
where $\sigma_{i}$ is the $i$th observational uncertainty and 
$\hat{ \bm{m}}$ is the model prior \rev{that is the uniform mean albedo map estimated from the observation 
\citep{2011ApJ...739L..62K}.} The regularization parameter $\lambda$ is arbitrary, and it balances the 
observational noise and the spatial resolution; the large value of $\lambda$ is likely to return the 
uniform map similar to the prior map. The above equation can be analytically solved in the following form,
\begin{align}
\bm{m}_{\rm est, \lambda} &= V\Sigma_{\lambda}U^{T}(\bm{\tilde{d}} - \tilde{G}\bm{\hat{m}} ) + \hat{\bm{m}}, \label{tikhonov_solve} \\
(\Sigma_{\rm \lambda})_{i,j} &= \frac{\kappa_{i}}{\kappa_{i}^{2} + \lambda^{2}}\delta_{i,j},
\end{align}
where $\tilde{d}_{i} = d_{i}/\sigma_{i}$, and $\tilde{G}_{i} = G_{i}/\sigma_{i}$. 
The matrices $V$ and $U$ are given by the singular value decomposition of $\bm{G} = U\Lambda V^{T}$, 
and $\kappa_{i}$ is the $i$th eigenvalue of the diagonal matrix $\Lambda$. 

\rev{The minimization with the Tikhonov regularization is equivalent to 
finding the maximum posterior probability in the Baysesian statistics, where 
the prior is imposed as the Gaussian-type function with the mean of $\hat{m}$ and the covariance matrix 
$\lambda^{-2} \delta_{i,j}$. The detailed discussion is 
shown in Appendix C in \cite{2012ApJ...755..101F}. }

\subsubsection{Sparse modeling}
Alternatively to Tikhonov regularization, we consider 
sparse modeling, which involves the combination of the L1-norm and Total Squared Variation (TSV) introduced in \cite{2018ApJ...858...56K} as 
the regularization terms for mapping planets. Then, the loss function $Q_{l1, {\rm tsv}}$ is defined as 
\begin{equation}
Q_{l1, {\rm tsv}} \equiv  \chi^{2} + \Lambda_{l} Q_{l} + \Lambda_{\rm tsv} Q_{\rm tsv}.  \label{eq:Q_l1_tsv} 
\end{equation}
The loss function $Q_{l1, {\rm tsv}}$ is composed of the chi-squared value, the 
L1-norm of the map $Q_{l}$, and TSV term  $Q_{\rm tsv}$ defined as follows:
\begin{align}
Q_{l} &\equiv \sum_{i}^{N_{\rm pixel}}|m_{i}|,  \\
 Q_{\rm tsv} &\equiv \sum_{i}^{N_{\rm pixel}}
\sum_{j}^{N_{\rm pixel}}\frac{1}{2} W_{i, j} (m_{i} -m_{j})^{2}, 
\end{align}
where $W_{i, j}$ is the neighboring matrix defined as 
\[
    W_{ i, j} = \begin{cases}
     1 & \text{if $i$-th and $j$-th pixels are adjacent.}\;\; \\
     0 & \text{otherwise.}
  \end{cases}
\]
In Eq (\ref{eq:Q_l1_tsv}), $\Lambda_{l}$ and $\Lambda_{\rm tsv}$ are 
the regularization parameters of L1 and TSV, respectively. The second term $\Lambda_{l} Q_{l}$ 
describes the sparsity of the solution. The larger value of 
$\Lambda_{l}$ gives more zero-valued pixels in the solution. The third term 
$\Lambda_{\rm tsv} Q_{\rm tsv}$ in Eq (\ref{eq:Q_l1_tsv}) is 
defined as the sum of the difference in the values of adjacent pixels. 
This term determines the smoothness of the solution. 
We solved the minimization of $Q_{l1, {\rm tsv}}$
with a monotonic variant of the fast iterative shrinking threshing 
algorithm (MFISTA; \cite{beck2009fast,beck2009fast2}), following 
\cite{2017AJ....153..159A} and \cite{2018ApJ...858...56K}. 

\rev{For the fair comparison with the Tikhonov regularization and its solution in Eq (\ref{tikhonov_solve}), we do not adopt non-negative condition for 
$m$, but such constrain would help to resolve the surface. 
One minimization for the data in Sec \ref{mock_sim} takes a few minute to finish, and the computational cost is scaled as $\mathcal{O}$(max($N_{\rm pixel}^{2}$, $N_{\rm pixel}N_{\rm data})$). 
The minimization with the L1 norm and TSV term is also same as 
finding the maximum posterior probability, where 
the prior is imposed as the combination of Laplace distribution for L1 norm and the Gaussian-type function for TSV. }

\subsection{Choices of regularization parameters in inverse modeling}
\rev{The optimization of regularization parameters is important 
issues in statistical methods, but the general discussion for the selection is difficult. 
In the global mapping problem with the Tikhonov regularization, \cite{2011ApJ...739L..62K} 
proposed to use $l$-curve method, which determines the optimal solution as the 
maximum curvature point of the model norm versus residuals corresponding to 
$\chi^{2}$ \citep{hansen2010discrete}. 
Here, $l$-curve method searches for the point with the maximum curvature between 
$\log  |\tilde{d} - \tilde{G}m_{\rm est, \lambda}|^{2}$ and $\log |m_{\rm est, \lambda} - \hat{m}|$, and  
the detailed description of the method is found in Appendix E in \cite{2012ApJ...755..101F}. }

\rev{In case of the sparse modeling with the combination of L1-norm and TSV, however, 
$l$-curve method cannot be applied straightforwardly because there is one additional 
regularization term. One might be able to extend $l$-curve to the higher-order method, 
but this is beyond the current scope of this work as there is no such previous study to our knowledge. Another possibility is the cross validation method, 
where the data are split into training and validation data, and the trained model 
is evaluated against the test data. However, this method favors the overfitted 
solution in the global mapping as discussed in Appendix \ref{sec:compare}. }

\rev{One possible strategy is to train regularization parameters by simulating several 
configurations, and apply the trained optimal 
regularization parameter to the real data \citep[e.g.][]{2019ApJ...875L...4E}. In the case of 
global mapping, we firstly prepare different
land distributions, simulate the observational data for the models, recover the maps from 
the simulated data, and choose the acceptable regularization parameters by 
comparing the recovered maps with the ground-truth images. 
Finally, one will recover the map from the real
data by adopting the regularization parameters, which are determined by the 
inject and recovery tests in the above. }

\rev{Given this procedure in mind, we focus on studying the potential of the sparse modeling in this paper; specifically, 
we attempt to investigate whether there exist possible combinations of regularization 
parameters in sparse modeling that recovers the better map than the Tikhonov regularization. 
For that, we compute a weighted residual sum of squares (WRSS) between the recovered maps and the ground map 
in order to identify the best map from the observation: 
\begin{equation}
{\rm WRSS} = \sum_{i}^{N_{\rm pixel}}  \bar{G}_{i} (m_{ {\rm est}, i} - m_{ {\rm true}, i})^{2},  \label{wrss}
\end{equation}
where the $\bar{G}_{i}$ is the time-averaged weight at $i$-th pixel, 
$m_{ {\rm est}, i}$ is the estimated model for the map, and $m_{ {\rm true}, i}$ is the 
true map. For regions without much visibility,  the term $\bar{W}_{i}$ suppresses the degree of discrepancy 
between the recovered and true maps in choosing the optimal map. }

\section{Mapping the cloudless Earth \label{mock_sim}}
\begin{figure}[h]
\begin{center}
 \includegraphics[width =16cm]{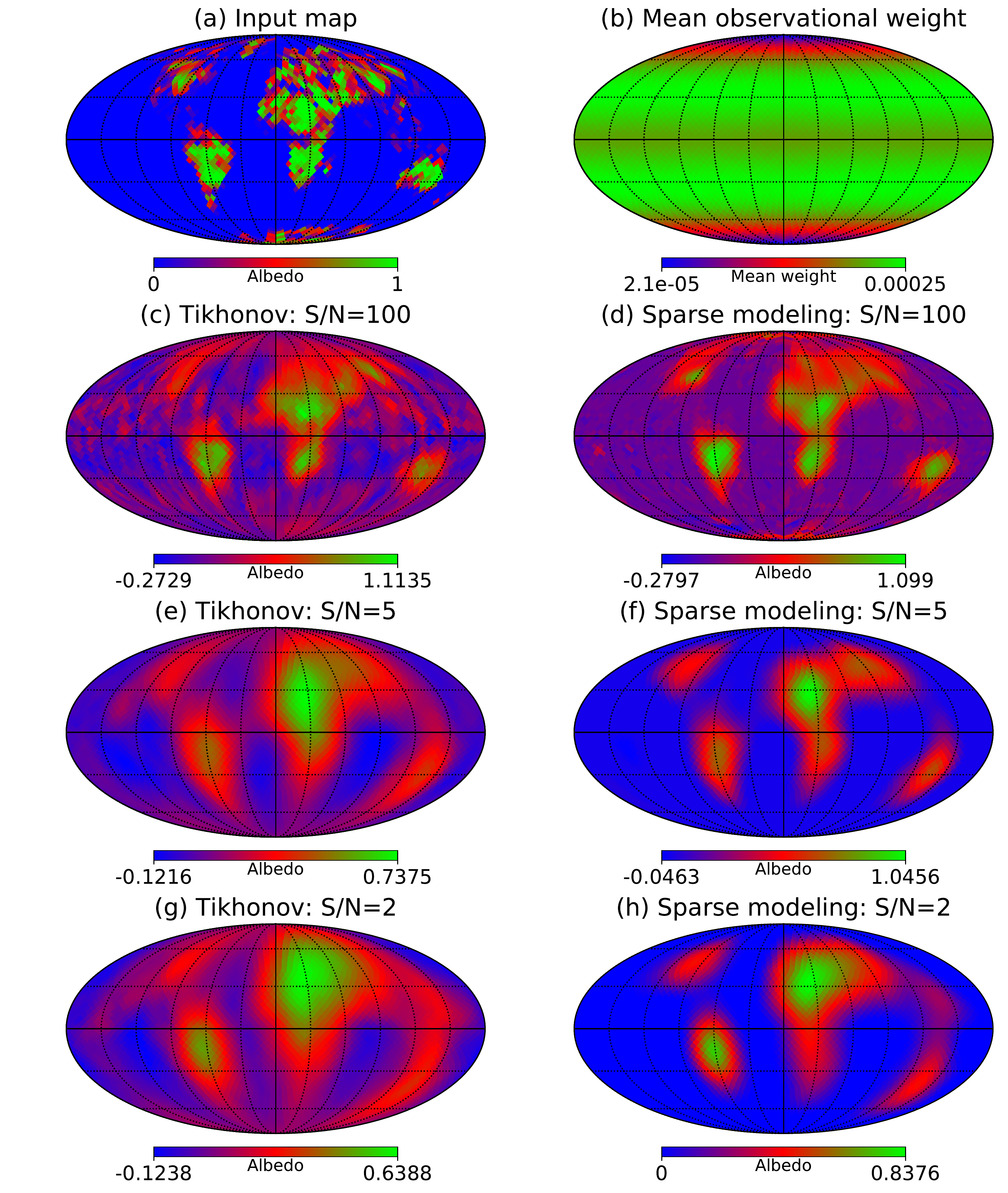}
 \caption{Mock albedo map of the Earth and the recovered surface estimated from 
 the lightcurves with S/N=2, 5.  (a) 
 Injected albedo map of the Earth. (b) Annual mean of the observational 
 weights $G_{i,j}$ of the mock data. (c) Recovered map based on 
 Tikhonov regularization (S/N=100). (d) Recovered map with S/N=5 based on regularization of the 
 L1-term an TSV. 
 (e) Same as (c), but for S/N=5. 
 (f) Same as (d), but for S/N=5. 
 (g) Same as (c), but for S/N=2. 
 (h) Same as (d), but for S/N=2. \label{fig:map_mock}}
\end{center}
\end{figure}

As a test bed, we adopt a static cloud-subtracted Earth model, as used in 
\cite{2016ApJ...822..112K}. Figure \ref{fig:map_mock}(a) shows this 
injected albedo map of the Earth after removal of the cloud-cover fraction with ISCCP D1 data (the cloud map of 2008 Jun 30 21:00). On this map, the ocean has zero albedo and land has a constant albedo after subtracting the cloud coverage. 
The spherical pixelization was realized using 
Hierarchical Equal Area Iso Latitude pixelation of the sphere (HEALPix) \citep{2005ApJ...622..759G} with 3072 pixels in total. With regard to the TSV terms, 
we calculated $W_{i,j}$, referring to the orders of pixels on the sphere. 
Concerning the geometry and orbital parameters, 
we assumed $i_{\rm inc}=0^{\circ}$, $\zeta=90^{\circ}$, $\Theta_{\rm eq}=180^{\circ}$, 
$P_{\rm spin}=23.93447$ hours, and $P_{\rm orb}=365.24219$ days.
We generated a one-year light curve with 1024 points at 
$\simeq$8 hours interval, and we added Poisson noise to the 
light curves by varying S/N$=2, 5, 100$. In the analysis, we adopted 
``Tikhonov" and ``L1-norm+TSV" regularization terms respectively for comparison. 
Figure \ref{fig:map_mock}(b) shows the mean weight of $G_{i, j}$ over the 
Earth surface in this mock observation. The Earth surface 
was globally surveyed except for the regions very close to the North and South poles.

Figure \ref{fig:map_mock}(c)-(f) shows the recovered maps estimated 
from the light curves with S/N=2, 5, and \rev{100} with the ``Tikhonov" and 
``L1-norm+TSV" regularization terms.  
The chosen regularization parameters are \rev{$\lambda=10^{0.1}$} for S/N=2, 
\rev{$\lambda=10^{0.3}$} for S/N=5, and \rev{$\lambda=10^{0.3}$ for S/N=100} 
according to the $l$-curve criterion \citep{hansen2010discrete} 
for Tikhonov regularization. For ``L1-norm+TSV" 
regularization,  we set \rev{$(\Lambda_{l},\Lambda_{\rm tsv})  = (10^{-0.50},10^{-0.25})$ for S/N=2 and  $(\Lambda_{l},\Lambda_{\rm tsv})  = (10\revnn{^{-0.25}},10^{-0.25})$ for S/N=5, 
and $(\Lambda_{l},\Lambda_{\rm tsv})  = (10^{0.75},10^{0.25})$ for S/N=100 as the optimal solutions by finding the minimum values of WRSS.}

\rev{Both methods generally succeed in recovering 
the major continents. In case of S/N = 100, both methods give well resolved 
maps, there is no large difference between them, although the sparse modeling 
still gives the better fitting.} On the other hand, as is evident from the 
comparison with the input map, Figure \ref{fig:map_mock}(d,f) 
shows better resolved and more consistent maps than Figure 
\ref{fig:map_mock}(c,e); the detailed structures of the 
continents (e.g., the shapes of South America and Africa) 
are well reproduced in Figure \ref{fig:map_mock}(d,f).
This tendency is more clearly seen in the comparison with S/N=2.  
The Tikhonov regularization (Figure \ref{fig:map_mock}(e)) 
fails to discriminate North and South America or the the Eurasian Continent and the 
Australian Continent. In contrast, sparse modeling (Figure \ref{fig:map_mock}(f)) successfully distinguishes 
these continents with well characterized coastlines. 
We note that the smaller $\lambda$ value gives a higher resolution with 
Tikhonov regularization, albeit with an over-fitted inferred map 
and induced noise. 

\begin{figure}[h]
\begin{center}
 \includegraphics[width =16cm]{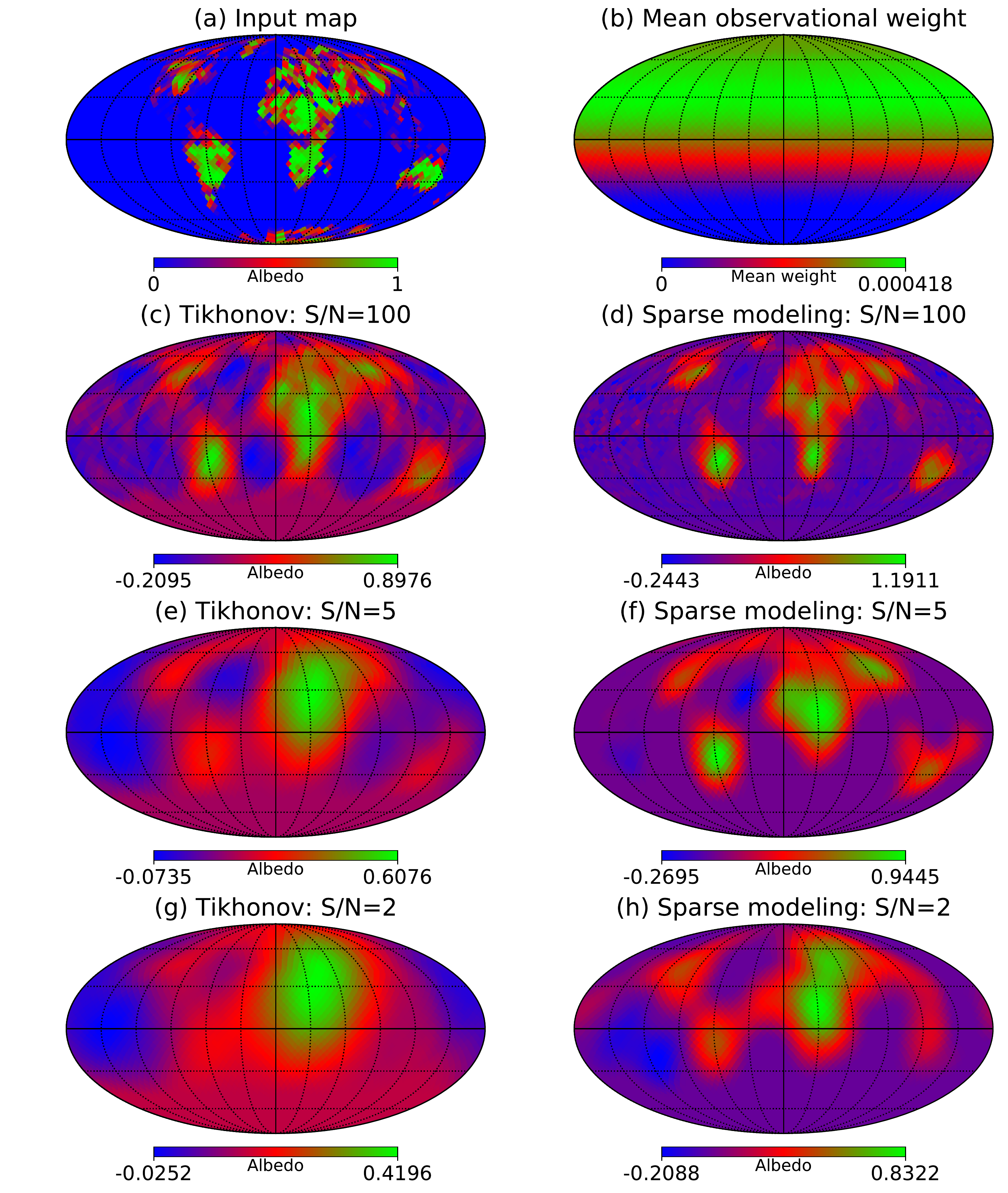}
 \caption{Same as Fig. \ref{fig:map_mock} but for $\zeta = 23.445^{\circ}$ 
 and $i=45^{\circ}$. \label{fig:map_mock_Earth}}
\end{center}
\end{figure}

\rev{In addition to the case of $i_{\rm inc}=0^{\circ}$ and $\zeta=90^{\circ}$, we also attempt $i_{\rm inc}=45^{\circ}$ and $\zeta=23.4^{\circ}$ to 
estimate the surfaces by exploiting simulated light curves with S/N=2, 5, and 100. We find that $\lambda = 10^{0.429}$ returns 
the optimal maps determined by the $l$-curve method for each case in the Tikhonov regularization. For sparse modeling, 
$(\Lambda_{l},\Lambda_{\rm tsv})  = (10^{-1.25},10^{-0.50})$ in 
the case S/N=2 and  $(\Lambda_{l},\Lambda_{\rm tsv})  = (10^{-1.00},10^{-0.50})$ in the case S/N=5, 
and $(\Lambda_{l},\Lambda_{\rm tsv})  = (10^{0.25},10^{-0.25})$ in 
the case S/N=100. Figure \ref{fig:map_mock_Earth} shows the 
recovered optimal maps obtained from minimizing WRSS for each case. In this setup, the north region of the Earth is 
preferably surveyed, and most of the south part is invisible due to the geometry. The sparse modeling successfully resolves 
the continents, especially in case of S/N=2 and 5. This example shows that the sparse modeling can work in different geometrical configurations. }

Differences among the reconstructed maps originate from various aspects. 
Tikhonov regularization is not physically motivated by the nature of the planetary surface. 
This term acts as a regulator for observational noise and the spatial resolution of 
the surface \citep{2011ApJ...739L..62K}. In contrast, the L1-norm efficiently 
identifies ocean regions because their albedo is zero. In addition, the TSV term suppresses the 
emergence of the bumpy structures on the maps, and the recovered maps become 
smoothed as a result of the minimization.

\clearpage

\newpage
\section{Application to observed light curves}
\begin{figure}[h]
\begin{center}
 \includegraphics[width =16cm]{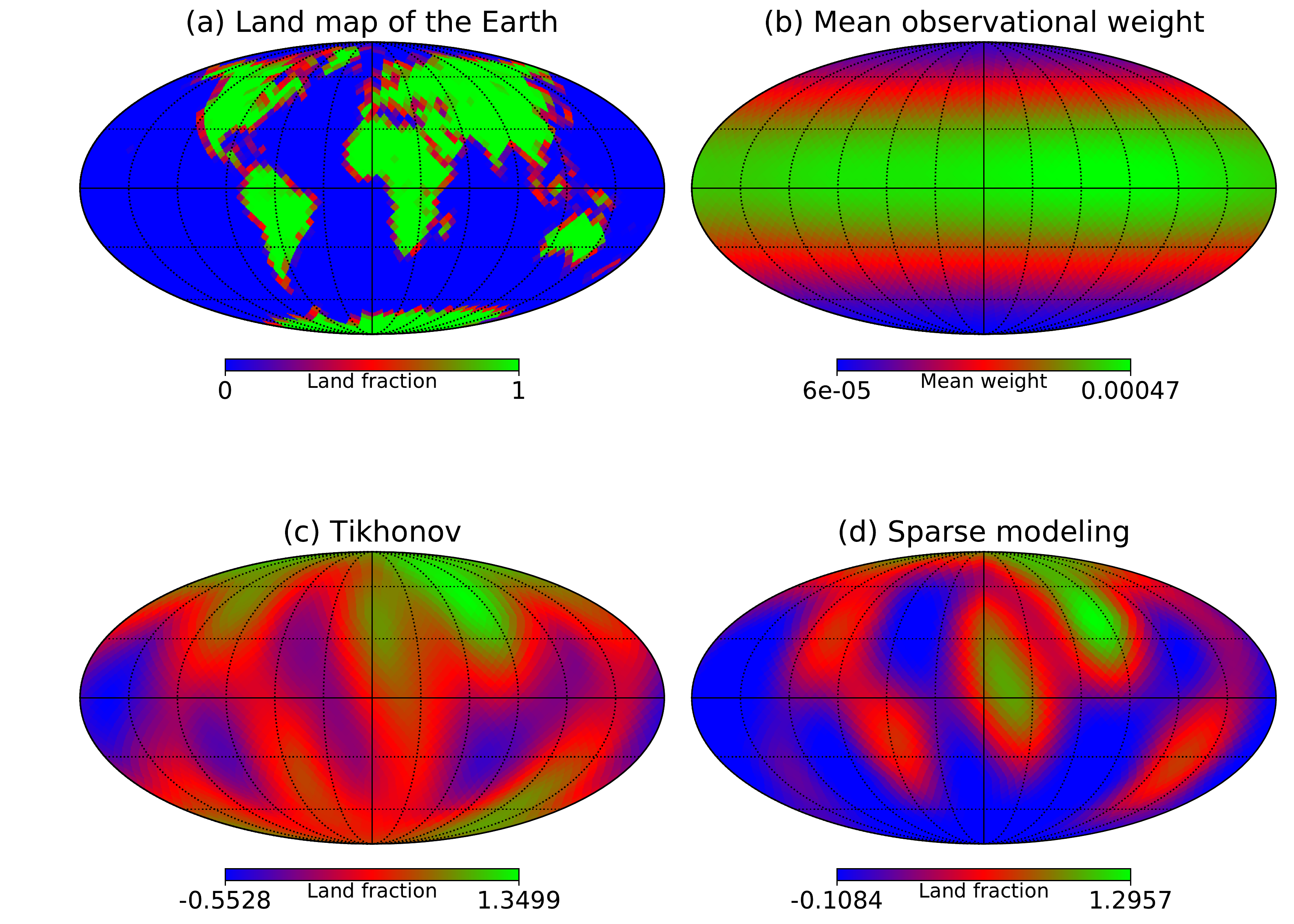}
 \caption{Mapping the surface of the Earth using 
 the 2-year DSCOVR/EPIC observations. (a) 
 Ground truth of pixelated land fraction surface map of the Earth \citep{2019ApJ...882L...1F}. 
 (b) Annual mean of the observational 
 weights $G_{i,j}$ in the 2-year DSCOVR/EPIC observations. (c) Recovered map with Tikhonov regularization 
 derived by \cite{2019ApJ...882L...1F}. (d) Recovered map with regularization of 
 L1-norm and TSV. The \rev{maps recovered from PC2 are converted into the values corresponding to the land fraction.}  \label{fig:fan_summary}}
\end{center}
\end{figure}

Recently, \cite{2019ApJ...882L...1F} reproduced the surface map of the 
Earth using real light curve observations of 
$\sim$10,000 DSCOVR/EPIC frames collected over a two-year period. 
\rev{Since the DSCOVR spacecraft is located at the first Sun-Earth Lagrange point (L1), the current configuration corresponds to $(i_{\rm inc}, \zeta) = (90^{\circ}, 23.4^{\circ})$ with varying $\Theta_{\rm eq}$ for $e_{\rm O}$ in Eq (\ref{eq:obs}) to be parallel from the Sun to the Earth.}  Observations were taken in 10 optical narrow band channels, and the principal components (PCs) were calculated among all the light curves to extract the surface feature. \rev{By exploiting Gradient Boosting Decision Tree, \cite{2019ApJ...882L...1F} found that the second strongest principal component (PC2) traces the surface inhomogeneity of planets. 
Particularly, \cite{2019ApJ...882L...1F} demonstrated that the time series of PC2 are linearly correlated with those of the overall land fraction, which is the summation of the land fraction weighted by $G$ viewed from the observatory at each phase. Here, the land fraction is taken from 
the Global Self-consistent, Hierarchical, High-resolution Geography Database \citep{1996JGR...101.8741W}, and it is shown in 
Figure \ref{fig:fan_summary} (a). In \cite{2019ApJ...882L...1F}, they 
recovered the land map from PC2 using the Tikhonov regularization. }

\rev{We analyze the same data as used in \cite{2019ApJ...882L...1F}. We 
check that the our analysis with the singular value decomposition analysis of multi-bands observations give 
the same PC2 as presented in \cite{2019ApJ...882L...1F}. 
For consistency, we adopt their normalized weight matrix, where the summation of weight at one epoch is equal to one, slightly 
different from the current definition of $\bm{G}$.} 
Figures \ref{fig:fan_summary}(b)-(c) show the mean weight matrix of the observation obtained
by the Earth Polychromatic Imaging Camera (EPIC) onboard DSCOVR 
and the corresponding recovered map, respectively. 
This observation is fairly insensitive to the north and 
south poles because DSCOVR is always located near the first Lagrangian point between the Earth and 
the Sun. In their solution, they adopted $\lambda = 10^{-1.5}$ as the regularization parameter in Eq (\ref{Eq:tikhonov}). 
Their estimation (Figure \ref{fig:fan_summary}(c)) 
captured the coarse features of the Earth surface, but not all the
continents were successfully recovered 
(e.g., South America and Australia). 

 For comparison, we solved the same problem using the L1-norm and TSV regularization terms. 
The light curves have negative and positive values owing to the nature of the principal 
component. In the solution derived by \cite{2019ApJ...882L...1F}, the regions with negative 
PC2 are more likely to be ocean. To make efficient use of sparsity, i.e., to associate zero values to ocean regions, 
we produced the non-negative light curves by subtracting the minimum value of the light curves from the overall light curves, and slightly offsetting the entire data set by 0.005. 
This ad hoc operation, or the choice of the added constant, does not significantly affect the results because the
constant offset in the light curve would only result in a constant change to the whole recovered map. 
Figure \ref{fig:fan_summary}(d) shows the recovered map, based on the L1-norm and TSV regularization. 
\rev{For the clear comparison, we convert PC2 of recovered maps into land fraction the corresponding 
by using the relation, PC2 $=$ 0.0753$\times$land fraction$-$0.0214, which is derived from the 
linear regression. In the minimization, we assume the associated noise $\sigma_{i} =1$ for simplicity. 
We adopted ($\Lambda_{l}, \Lambda_{\rm tsv}) = (10^{-3}, 10^{-3})$ in Eq (\ref{eq:Q_l1_tsv}) as the optimal solution chosen 
by minimizing WRSS between the recovered map and the land fraction of the Earth in Figure \ref{fig:fan_summary}(a). } 
Notably, the mapping by the new regularization term resolves the
structure of South America and Australia, 
which are blurred and connected to Antarctica in Figure \ref{fig:fan_summary}(c). 
It also successfully resolves other continents very consistently with the features 
of the Earth surface. However, our newly proposed map does not recover 
Antarctica as implied in Figure \ref{fig:fan_summary} \rev{(c)}.
This may be due to its low observational weights (Figure \ref{fig:fan_summary}(b)) and different spectral features than other continents. 

\begin{figure}[h]
\begin{center}
  \includegraphics[width =18cm]{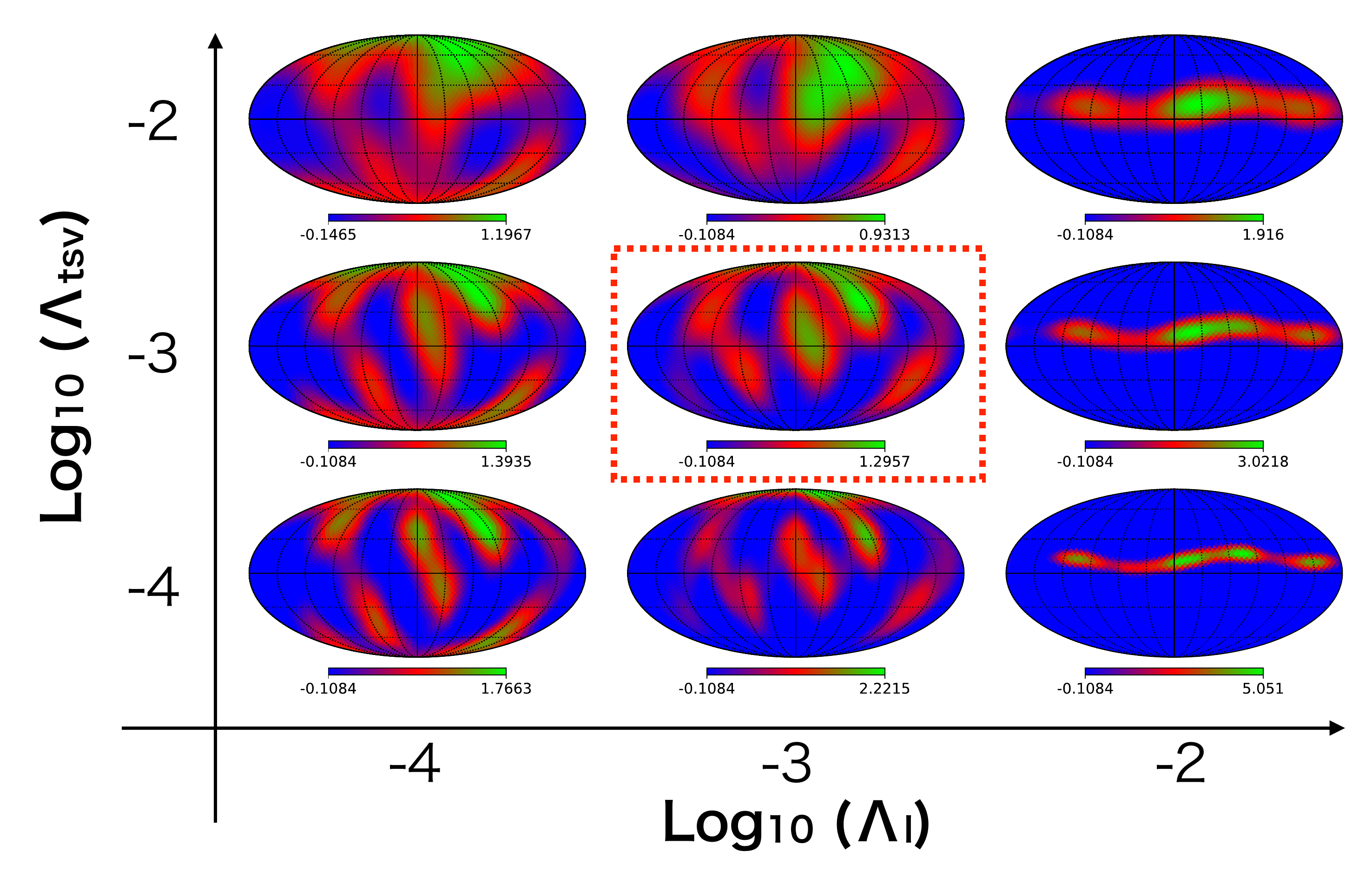}
 \caption{Recovered surface maps with different combinations of $\Lambda_{l} = (10^{-4},10^{-3},10^{-2})$ and $\Lambda_{\rm tsv} = (10^{-4},10^{-3},10^{-2})$. The optimal solution is shown in the 
 center of the panel being surrounded by red dotted lines.  \label{fig:fan_sparse_comp}}
\end{center}
\end{figure}

Finally, we show solutions with 9 different combinations of 
$\Lambda_{l} = (10^{-4},10^{-3},10^{-2})$ and $\Lambda_{\rm tsv} = (10^{-4},10^{-3},10^{-2})$ in Figure \ref{fig:fan_sparse_comp}. Toward the larger value of $\Lambda_{l}$, the reconstructed maps become more sparser, and inconsistent with the ground truth. On the other hand, with smaller value of $\Lambda_{l}$, the solutions become more similar to those obtained from the Tikhonov regularization (Figure \ref{fig:fan_summary}(c)). The optimal solution with $(\Lambda_{l}, \Lambda_{\rm tsv}) = (10^{-3}, 10^{-3})$ is exactly in the middle of these two kinds of solutions, and 
the three cost functions in $Q_{l1, {\rm tsv}}$ balance each other at this point.

\section{Demonstration of global mapping in future observation } 

In the future space direct imaging observations (e.g. the Habitable Exoplanet Observatory (HabEx) and the Large UV/Optical/IR Surveyor (LUVOIR)), it would be 
unrealistic to have continuous occupancy of an instrument for years. In addition, the instrumental and astrophysical noises in actual exoplanet observations would be orders of magnitude larger than those in Earth's light curves. It is necessary to take the consideration of much less time frames and higher noise level for exoplanet observations. In this section, we test the feasibility of the method under a more realistic assumption. 

As a fiducial system, we consider an Earth size planet around 
a Sun-like star with T = 5780 K $(T_{\odot})$ and $R_{\star} = R_{\odot}$ 
at a distance of \revn{5} pc. The semimajor axis of
the planetary orbit is assumed to be 1 AU, and the 
planetary flux in each band is calculated by
convolving the planet's reflectivity with the stellar spectrum. For 
simplicity, we assume the reflection to be
Lambertian, and the phase angle $\alpha$ is 
asssumed to be $90^{\circ}$, where the planet is at a 
quadrature. We consider a \revnn{LUVOIR-like} telescope with a diameter $D=\revn{8}$ m. The coronagraph
design contrast is assumed to be $10^{-10}$, and the 
end-to-end throughput in the coronagraph is taken to be 0.3 \footnote{\revnn{Expected coronagraph contrast and end-to-end throughput of the ECLIPS instrument is adopted from the LUVOIR Final Report (http://www.latex-cmd.com/struct/footnote.html). The average optical throughput in the visible band is assumed to be 0.277 in the imaging mode.}}. 
The integration time is 1.8 hours, which corresponds to the average time interval in the 
current observation. For simplicity, we consider the snapshot of each image 
rather than the smeared image with 1.8 hours integration. 
\rev{However, if we take into account the smearing effect, the recovered image would be 
blurred by 27 degrees, which corresponds to 1.8 hours, around the spin axis. 
Although the current cadence is already limited by 
the observational strategy of DSCOVR, one will be able to evade this problem by 
adopting shorter cadence. }

We calculate the observational noises using {\tt coronagraph},
which is an open source Python package for computing the noise 
of space direct imaging missions 
\citep{2016PASP..128b5003R,2019JOSS....4.1387L}. 
Adopting the imaging mode in {\tt coronagraph}, we compute S/N in each band. 
As the DSCOVR observation uses narrow-band filters 
(317, 325, 340, 388, 443, 552, 680, 688, 764 and 779 nm), 
we reassign the band centers and full width at half 
maximum (FWHM) to emulate broadband filters of our mock observatory. For instance, we take the average of the light curves in three narrow-band filters of DSCOVR (317, 325, and 340 nm) as the light curve in a single band. In the 
similar manner, we combine the light curves in the filters with 
band centers of (680 and 688 nm) and 
(764 and 779 nm) by averaging their light curves. As a result, 
we have the light curve with six broadband filters with band centers = (325, 388, 443, 552, 684, 770 nm) and FWHM = (30, 33, 22, 87, 45, 41 nm). 

\begin{figure}[h]
\begin{center}
  \includegraphics[width=8cm]{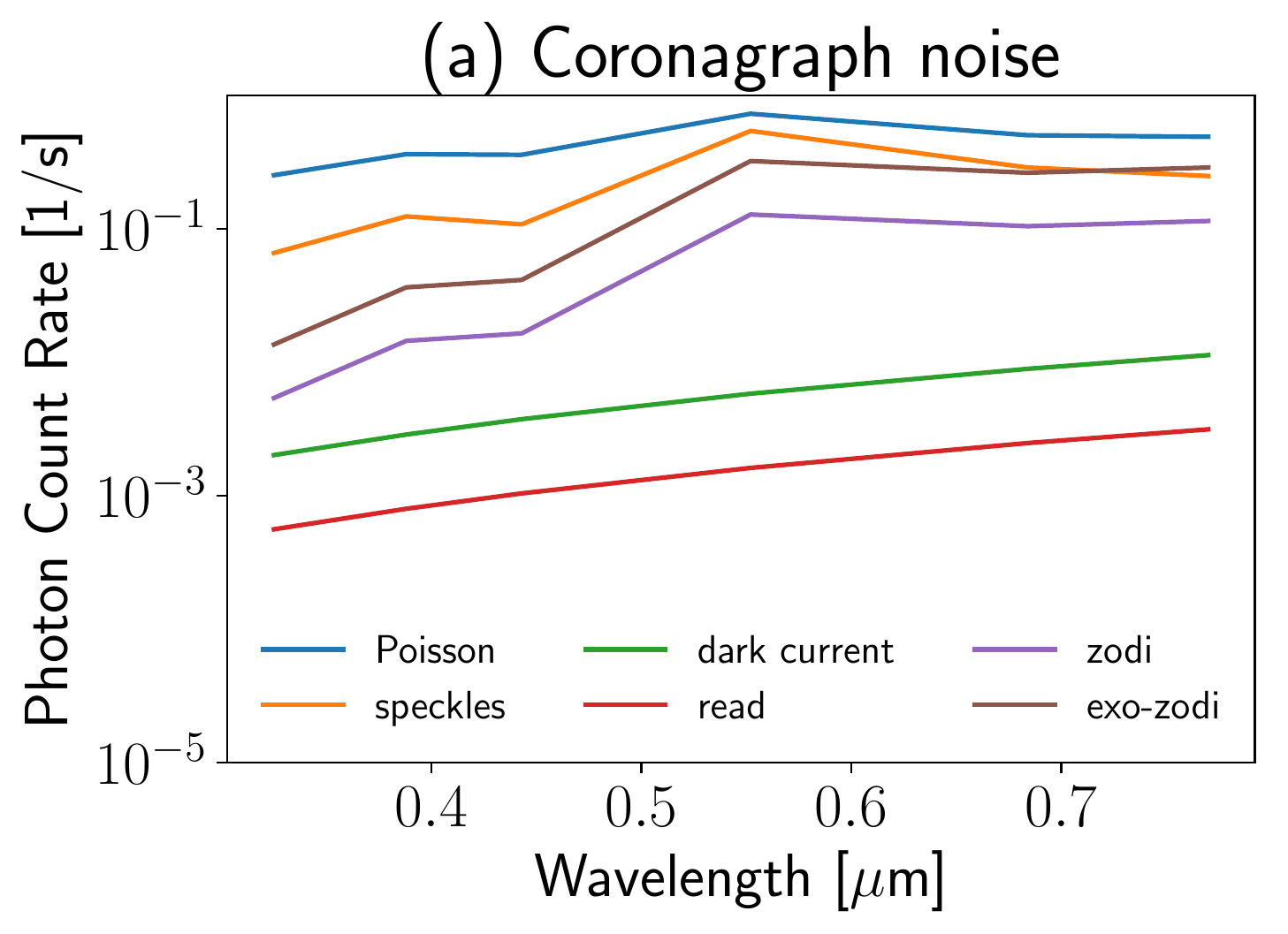}
    \includegraphics[width =9cm]{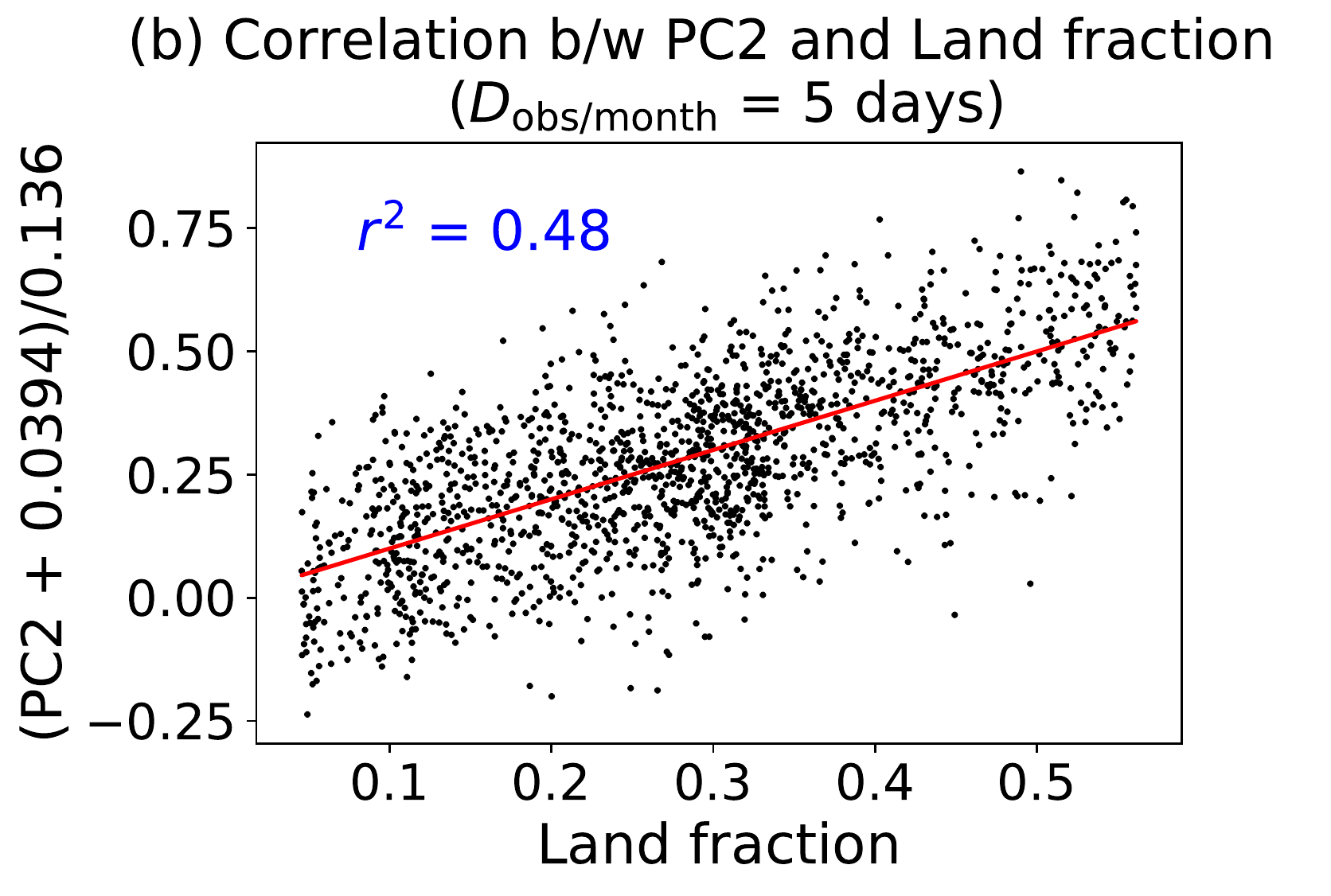}
 \caption{(a)Statistical noises in the unit of photon counts derived by {\tt 
 coronagraph} \citep{2016PASP..128b5003R,2019JOSS....4.1387L}. (b) 
 Scatter plot of land fraction and PC2 extracted from light curves 
 with $D_{\rm obs/month} = 5$ days. Red line shows a linear fitting to the data, and the coefficient of determination $r^{2}$ = \revn{0.48}. 
 \label{fig:noise_pdf}}
\end{center}
\end{figure}

Figure \ref{fig:noise_pdf} shows the compositions of the noise sources based on \cite{2016PASP..128b5003R}. The calculation is basically based on the sample code ``luvoir\_demo.py" in {\tt coronagraph}. The noise sources are composed of Poisson noise, local zodiacal light, exozodiacal light, dark current, read noise, and speckle noise. 
In the current case, the dominant noise sources are Poisson noise and \revn{speckle noise}. In the reconstructed broadband filters, we find 
S/N = \revn{(10.50, 15.17, 15.13, 26.92, 16.18, 15.40)} assuming 
1.8 hours integration. 

Instead of a continuous 2-year observation, 
we assume multi-epoch observations as those presented in \cite{2016MNRAS.457..926S} and \cite{2018AJ....156..146F}. 
We divide 1 year into 12 blocks ($\simeq$ 1 month), 
and use the first $D_{\rm obs / month} = 1$ or $5$ days in each block. 
Adopting S/N as above, 
we inject Gaussian noise to the light curves, and decompose them into principal components using \rev{SVD}. 
As revealed in \cite{2019ApJ...882L...1F}, the second strongest principal
component (PC2) is linearly correlated with the land fraction; the coefficient of determination $r^{2}$ is $0.91$
in their paper. Figure \ref{fig:noise_pdf} shows the scatter plot of PC2 and
the land fraction in case of $D_{\rm obs/month}=5$ days with $r^{2}=\revn{0.48}$,
which implies the weak correlation. Similarly, we find $r^{2} = \revn{0.48}$ for $D_{\rm obs/month}=1$ day and \rev{$r^{2} = \revn{0.50}$} for the full data \rev{set}. On the other hand, we find $r^{2} = 0.93$ without the injection of noise, so the noise would account 
for the weak correlation in the current case. 
We fit a linear function of the land fraction to the PC2, and \rev{we find 
\revn{PC2 $= 0.290\times$land fraction$- 0.083$} for $D_{\rm obs/month}=1$ day
\revn{PC2 $= 0.136\times$land fraction$- 0.039$} for $D_{\rm obs/month}=5$ days. The coefficients 
roughly differ by a factor of $\sqrt{5}$ between $D_{\rm obs/month}=1$ and $5$ days, and this is because 
the normalization of PC2 is equal to 1 by definition. Finally, we adopt 
\revn{(PC2 $+ 0.083$)/0.290} for $D_{\rm obs/month}=1$ day and \revn{(PC2 $+ 0.039$)/0.136} for
$D_{\rm obs/month}=5$ days for solving the map in order for $\bm{m}=0$ corresponds to zero land fraction i.e. ocean. }

\begin{figure}[h]
\begin{center}
  \includegraphics[width =16cm]{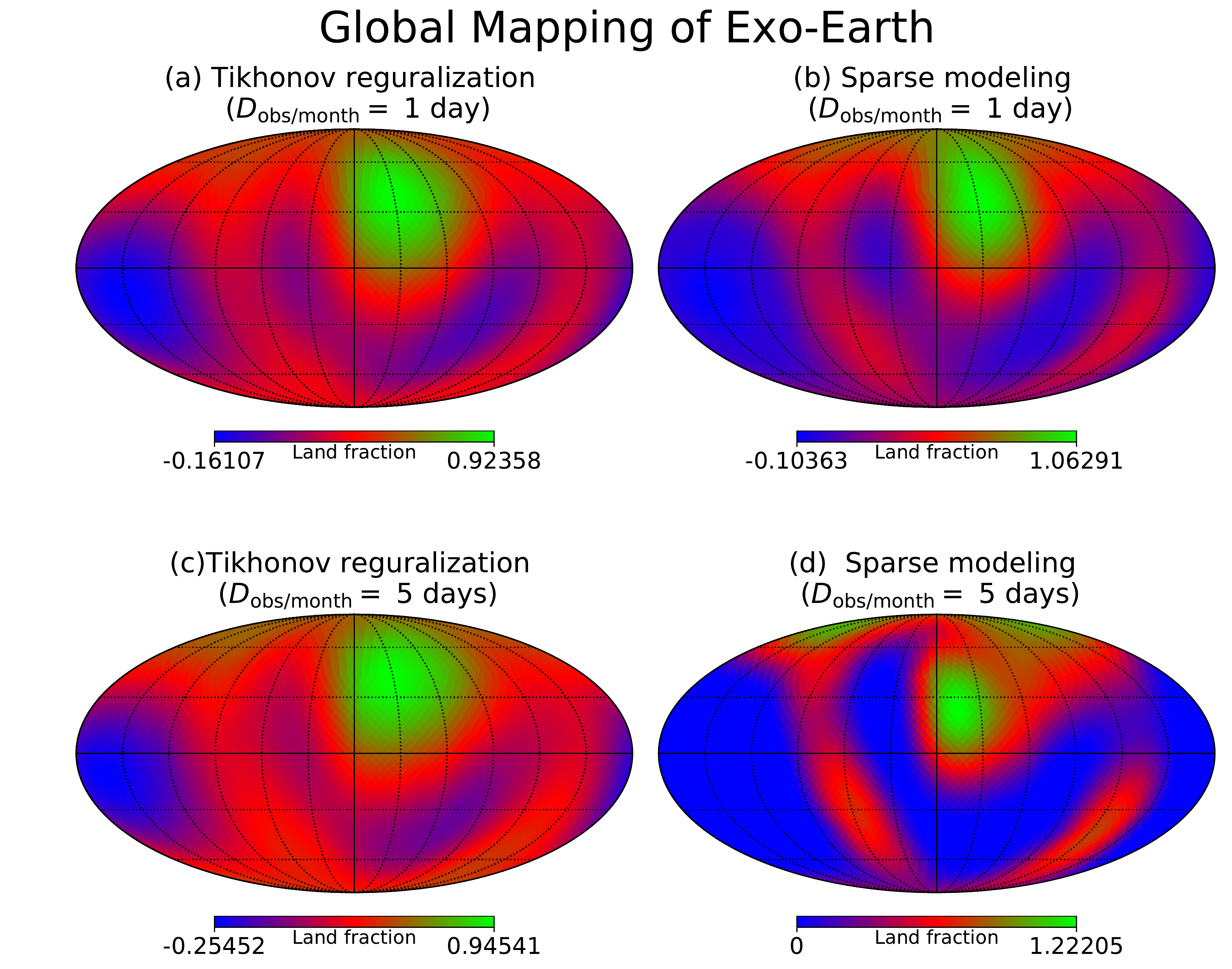}
 \caption{Global mapping of the DSCOVR data assuming observations from a distance of 5 pc using 
 Tikhonov regularization (a,c) and Sparse modeling (b,d). 
 \revn{The observational duration is one day in one month (panels (a) and (b); $D_{\rm obs / month} = 1$ day), 
 and five days (panels (c) and (d); $D_{\rm obs / month} = 5$ days). The recovered maps are chose by 
 minimizing WRSS between the recovered maps and the land map of the Earth. }
  \label{fig:global_map}}
\end{center}
\end{figure}

\rev{Figure \ref{fig:global_map} \revn{shows} the recovered surface maps using Tikhonov regularization and sparse modeling for $D_{\rm obs/month}=1$ and 5 days, respectively. We express the recovered maps in representation of land fraction by exploiting linear relation for PC2, and \revn{the optimal solutions are obtained by minimizing WRSS between the recovered maps and the land map of the Earth in Figure \ref{fig:fan_summary} (a) for both Tikhonov regularization and sparse modeling.} Notably, even in case of $D_{\rm obs / month} = 1$ day, the planetary surface is roughly resolved in the Tikhonov regularization. Moreover,  the sparse modeling (Figure \ref{fig:global_map} (b)) successfully resolves the Australia, Afro-Eurasia, North America, and South America \revn{although it is still blurred.} On the other hand, observations with $D_{\rm obs / month}$ of 5 days in Figure \ref{fig:global_map}} enable us to recover the maps with very similar accuracy to those produced from continuous observations. These results encourage the mapping of ``Second Earth" using future directly imaging missions such as HabEx and LUVOIR.

\section{Summary}
The use of reflected light curves for mapping surfaces is an important probe 
of exoplanet surface inhomogeneity. 
Previous studies of recovering surface maps from observed light curves basically regularized 
observational noise with Tikhonov regularization without the consideration of 
physical properties of planetary surfaces \citep[e.g][]{2011ApJ...739L..62K,2012ApJ...755..101F,2018AJ....156..146F,2019ApJ...887L..14B}. 
\rev{In addition to those studies with pixel discretization on the sphere, \cite{2019arXiv190312182L} exploited 
the spherical harmonics functions for solving the mapping problem using the TESS light curve of the Earth, 
and they adopted priors, which suppress high order features of the surface.} 
In this paper, we introduced sparse modeling (L1+TSV regularization) 
to estimate a planetary surface efficiently 
by exploiting the surface properties, (a) distinct spectral difference between land and ocean 
and (b) continuity of planetary surfaces. As a test calculation, we 
injected and recovered the mock albedo map of the Earth using 
both regularization terms, and found that sparse modeling 
reproduces more consistent continental surface distributions than \rev{the Tikhonov regularization. }
We also applied our method to observations of the Earth obtained by DSCOVR to investigate the advantage of sparse modeling 
in real data analysis. We found that sparse modeling 
successfully recovers the continents (Australia and South America) that are not clearly estimated by Tikhonov regularization. We also showed that 1 or 5 days per month in 2-year observation \revnn{would enable us} to retrieve the surface map of an Earth analog at a distance of \revn{5} pc.

In this paper, \rev{we} demonstrate that the choice of regularization terms 
in exoplanet mapping can significantly affect the resulting recovered maps. 
\rev{The regularization terms can be seen as the priors from a Bayesian viewpoint, 
and our study shows the importance of selecting regularizations or priors for finding the correct map from the observation. In this viewpoint, the discretization on the planetary surface rather than the spherical harmonics would be intuitive to put reasonable priors that exploit characteristics of planetary surfaces. We also attempt other types of regularization terms, and the combination of L1+TSV regularization is likely to be
the best choice among possible regularizations in case of the Earth. In the current sparse modeling, we make use of the sparsity 
coming from the very low albedo of the ocean, but even for a planet without the sea, we can apply the method by offsetting the data to force an albedo of a particular surface type to be zero. 
In this paper, we consider only the case of the Earth, and other} continental configurations, geometries, regularization terms, and optimal
choice of the regularization parameter for unknown surface are subject to future investigation.

The authors thank the DSCOVR team for making the data publicly available. 
We also thank Yasushi Suto and Shiro Ikeda for fruitful discussions on 
the sparse modeling and its application on the mapping, and 
thank Yuk L. Yung for discussions on surface map retrieval 
using DSCOVR observations.  This work is supported by JSPS KAKENHI 
Grant Numbers 14J07182 (M.A.), JP17K14246, and JP18H04577, 
and JP18H01247 (H.K.). M.A. is also supported by the Advanced Leading Graduate 
Course for Photon Science (ALPS) and by the JSPS fellowship. This work was also supported by the
JSPS Core-to-Core Program “Planet$^2$”.
\appendix

\section{Comparison of different regularizations \label{sec:other_reg}}
\rev{In this section, we compare six different regularizations in mapping problem. For that purpose, we assume the same condition as done in Section \ref{mock_sim}; S/N=5, $i_{\rm inc}=0^{\circ}$, and $\zeta = 90^{\circ}$. By comparing the recovered maps from simulated light curves, we demonstrate that the L1+TSV is likely to be the best regularization for the mapping problem. }

\rev{For the comparison, we introduce additional regularization, Total Variation (TV), defined as 
\begin{equation}
 Q_{\rm tv} \equiv \sum_{i}^{N_{\rm pixel}}
\sum_{j}^{N_{\rm pixel}}\frac{1}{2} W_{i, j} |m_{i} -m_{j}|, 
\end{equation}
where we exploit the same neighboring matrix $W_{i, j}$ as used for the TSV term.
We define the regularization parameter 
for TV as $\Lambda_{\rm TV}$. 
The above expression cannot be differentiable when $m_{i} = m_{j}$, 
so we use replace $|m_{i} -m_{j}|$ by $\sqrt{(m_{i} -m_{j})^{2} + \epsilon^{2}}$, where 
$\epsilon=10^{-8}$ in this study, when we compute the derivative. 
With TV term, we can search for solutions with sparsity in 
derivatives of the surface albedo in the current problem. Since the 
planetary surface is roughly divided in to several types with boundaries, 
the TV term possibly helps to identify the clear edges. }

\begin{figure}[h]
\begin{center}
 \includegraphics[width =16cm]{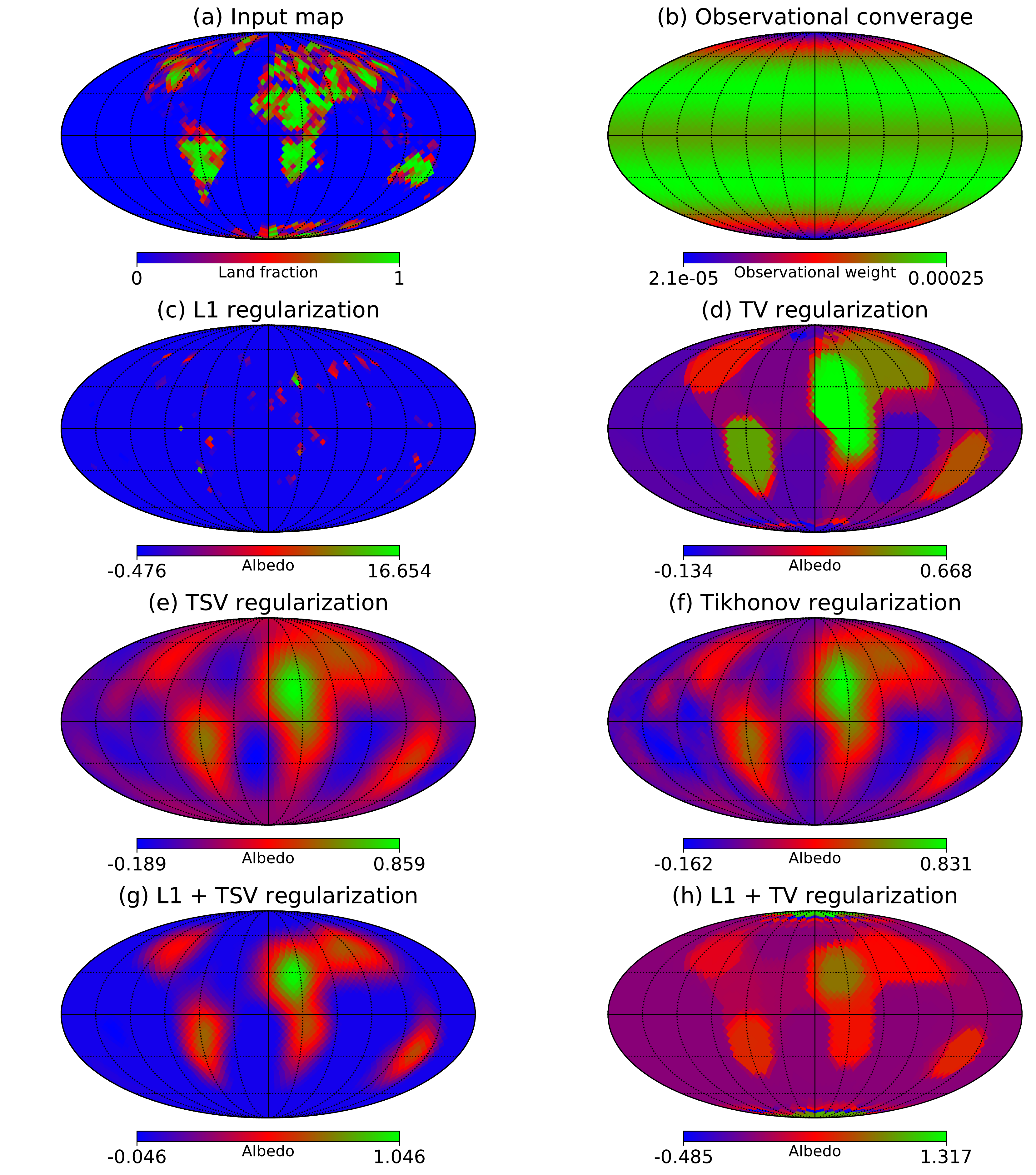}
 \caption{\rev{Comparison of recovered maps in different methods (S/N=5)(a) 
 Injected albedo map of the Earth. (b) Annual mean of the observational 
 weights $G_{i,j}$ of the mock data. (c) Recovered map based on 
L1 regularization. (d) Recovered map based on 
TV regularization.
 (e) Recovered map based on 
Tikhonov regularization. 
 (f) Recovered map based on 
L1+TSV regularization. 
 (g) Recovered map based on 
L1+TV regularization. }
 \label{fig:map_different_reg}}
\end{center}
\end{figure}

\rev{Using this new term, we exploit six different regularizations in total for estimating the planetary surface: 
L1+TSV, TV, L1 + TV, TSV, Tikhonov regularization, and L1 norm. To find the optimal solutions, 
we implement a grid search, where the ranges of the regularization parameters are from 
$10^{-3}$ to $10^{1}$. The optimal regularization parameters are 
$(\Lambda_{\rm l}, \Lambda_{\rm tsv})=(10^{-0.25},10^{-0.25})$ for  L1+TSV, 
$\Lambda_{\rm tv} = 10^{-0.5}$ for TV,  $(\Lambda_{\rm l}, \Lambda_{\rm tv})=(10^{0.56},10^{-0.33})$ for  L1+TV, 
$\Lambda_{\rm tsv} = 10^{0.25}$ for TSV, $\lambda = 10^{0.30}$ for Tikhonov regularization, 
and $\Lambda_{\rm l} = 10^{-0.5}$ for L1 norm. Figure \ref{fig:map_different_reg} shows the comparison of recovered maps with six 
different regularizations, where L1+TSV and Tikhonov regularization are already 
studied in Section \ref{mock_sim}. The maps are determined by minimizing 
WRSS in Eq (\ref{wrss}) from comparison with the input model of the Earth. 
Consistently with the expectation, TV term well separates the different regions by making 
clear boundaries between them, but it also tries to connect the continents 
by introducing new layers, which should be oceans in real, at the same time.
This is also the case for the combination of L1+TV. On the other hand, 
in case of the L1 norm alone, the anomalously high albedo values are associated with 
small pieces on the surface unlike the real distribution of the Earth surface. In the 
future observation of exoplanets in direct imaging, a planetary radius and albedo 
can be degenerate, and L1 norm would misunderstand the Exo-Earth as the 
very large planet, whose reflectivity mostly comes from small regions. 
In summary, we find WRSS = 0.0249, 0.0271, 0.0282, 0.0286, 0.0292, and 0.626 for L1+TSV, TV, L1 + TV, TSV, Tikhonov regularization, and 
L1 norm, respectively. Therefore, L1+TSV is the best choice among current possibilities. }

\section{Cross validation, l-curve method, and comparison with ground truth \label{sec:compare}}

\begin{figure}[h]
\begin{center}
 \includegraphics[width =16cm]{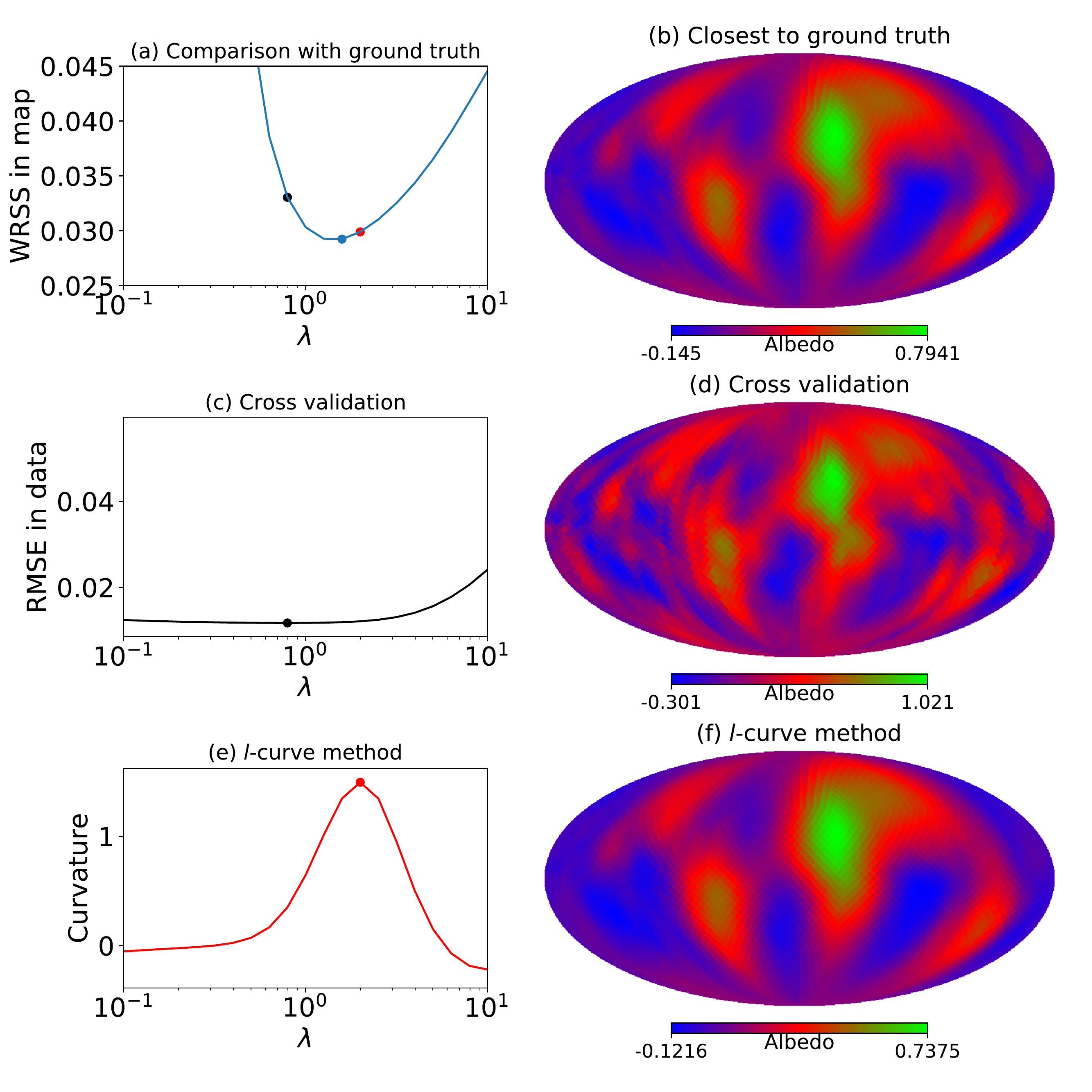}
  \caption{\rev{Comparison of recovered maps from the lightcurves with S/N=5 in Tikhonov regularization. Blue, black, and red points 
  correspond to optimal maps in panel (b), (d), and (f), respectively. 
 (a) Root mean squared errors (RMSE) between recovered maps and ground truth map. 
 (b) Recovered map obtained from comparison with ground truth map with $\lambda = 10^{0.2}$.
 (c) Root mean squared errors in cross validation. 
 (d) Recovered map obtained from the cross validation  with $\lambda = 10^{-0.1}$. 
 (e) Curvature exploited in $l$-curve method. 
 (f) Recovered map obtained from $l$-curve method with $\lambda = 10^{0.3}$. }
 \label{fig:map_comp}}
\end{center}
\end{figure}

\rev{In this section, we compare the cross validation and $l$-curve method for recovering the maps 
in the Tikhonov regularization. Here, the cross validation is a statistical method, in which 
we split the data into subsets, perform training using the data except for one test subset, and 
assess the goodness of the trained model by comparing its prediction with the remaining test data. 
In the current cross validation, we adopt 10-fold for splitting the data by the random selection, and 
compute the root mean squared errors by calculating the difference between 
the model light curves simulated from trained models and the rest of of the data for the test. 
In the analysis, we adopt the same light curve and geometrical configuration as used in case of S/N=5. 
The panel (a)  in Figure \ref{fig:map_comp} shows the deviations of recovered maps from the ground truth map 
depending on the regularization parameter $\lambda$. 
The blue point corresponds to the map closest to the ground truth map, the black point is obtained from 
the cross validation panel (c), and the red point is determined by the $l$-curve method in panel (e). 
The panels (b), (d), and (f) correspond to the recovered maps obtained from the 
comparison with the ground truth map, the cross validation, and $l$-curve method, respectively. }

\rev{The comparison demonstrates that the 
cross validation gives the overfitted map, and it is fairly deviated from the ground truth map. In addition, 
root mean squared errors in the cross validation are
significantly insensitive to $\lambda$, implying that the method can 
return the very overfitted solution with reasonable root mean squared errors. 
On the other hand, the $l$-curve method returns the map close to the best possible map by comparison with the ground truth map. 
Therefore, the $l$-curve method works at the same level of the comparison with the ground truth, although the cross validation does not. }

\bibliography{export-bibtex}
\bibliographystyle{aasjournal}

\end{document}